\documentstyle[12pt,aaspp4,flushrt]{article}
\begin{document}

\title{Dark Matter}

\author{Neta Bahcall\\
Princeton University\\Department of Astrophysical Sciences\\Princeton, NJ
08544-1001 USA}

Observations in the optical, in X-rays, and gravitational lensing of
galaxies, clusters of
galaxies, and large-scale structure are beginning to provide clues to the
dark matter problem.  I
review the impact of these observations on some of the main questions
relating to dark matter:  How
much dark matter is there? Where is it located? What is the nature of the
dark matter? and what is the
amount of baryonic dark matter.

\section{Introduction}
The evidence for the existence of large amounts of unseen ``dark matter''
in the universe, especially
in halos around luminous galaxies and in group and rich clusters of
galaxies, has been known for a
long time (see reviews by [\cite{FG:1979}]; [\cite{Trimble:1987}];
[\cite{Peebles:1993}]. The
presence of the dark matter component is inferred from the observed motions
(of stars, gas, and
galaxies) in and around galaxies and clusters of galaxies; from the
emission by hot gas in galaxy
halos and in groups and clusters; and from observations of gravitational
lensing by foreground
galaxies and clusters of galaxies.

While it is clear that only a small fraction of the mass of the universe is
in visible form --- the
rest being dark --- some of the most fundamental questions about dark
matter still remain open:

\begin{itemize}

\item How much dark matter is there in the universe? Is there enough matter
to close the universe (with
$\Omega_{m} \equiv \rho_{m}/\rho_{c} =3D 1$), or do we live in a low-density
universe ($\Omega_{m} < 1$)?

\item Where is the dark matter located? Is it mainly associated with the
light distribution (i.e., in
and around galaxies and clusters of galaxies), or is it mostly hidden in
the ``voids'', where no light
is concentrated?

\item What is the dark matter made of? Is it baryonic matter, or is it
composed mostly (or partly) of
non-conventional, non-baryonic, new type of matter such as cold (e.g.,
axions) or hot
(massive neutrinos) dark matter?

\item What limits can be placed on the amount of baryonic dark matter in
the universe?

\end{itemize}

I will address these questions and review the constraints placed by recent
observations.  While no
definitive answers are yet available to these questions, some hints are
emerging.  Observations over
the next decade are likely to yield some solutions to these fundamental
questions.

\section{Where is the Dark Matter?}

A large amount of dark matter exists in halos around luminous galaxies and
in groups and
rich clusters of galaxies [\cite{Ostriker:1974}]; [\cite{Einasto:1974}];
[\cite{FG:1979}],
[\cite{Davis:1980}]; [\cite{Trimble:1987}]. The presence of the dark matter
component is inferred
from the high rotation speed of gas and stars in the outer parts of spiral
galaxies [\cite{FG:1979}];
[\cite{Trimble:1987}], the high-velocity dispersion and extended X-ray halo
of hot gas in elliptical
galaxies  [\cite{FG:1979}]; [\cite{Lauer:1985}]; [\cite{Forman:1985}];
[\cite{Trimble:1987}];
[\cite{Fabbiano:1989}]; [\cite{Mushotzky:1994}], and the high-velocity
dispersion and gas
temperature in clusters of galaxies [\cite{FG:1979}]; [\cite{Trimble:1987}];
[\cite{Lubin:1993}].  More recent observations using gravitational lensing
as direct mass
estimators of galaxies and clusters further support these findings
[\cite{Kaiser:1993}];
[\cite{Tyson:1996}]; [\cite{Brainerd:1996}].  The relative contribution of
the dark matter
component is usually specified in terms of the mass-to-light ratio,
$M/L$; this reflects the total amount of mass relative to the total light
within a given scale.  It is
well known [\cite{Ostriker:1974}]; [\cite{Davis:1980}];
[\cite{Blumenthal:1984}];
[\cite{Rubin:1993}] that, on average, the $M/L$ ratio increases from the
bright, luminous parts of
galaxies to their faint halos, with further increase to systems with larger
scales such as groups and
rich clusters of galaxies. This increase suggests that there is relatively=
 more dark than luminous
matter with increasing scale.  This has led to the general belief that
clusters have more dark matter
per unit luminosity than individual galaxies and that superclusters may
have even more. This widely
accepted monotonic increase of $M/L$ with scale determines to a large
extent the prevalent views about
the location of the dark matter and the total mass density of the universe.
The universal mass density
can be estimated from the observed mean luminosity density in the universe
when multiplied by the $M/L$
ratio observed on large scales. On the scale of clusters ($\sim 1 $ Mpc),
this method suggests a total
mass density of $\Omega_{m} \sim 0.2.$ If $M/L$ continues to increase on
large scales, however, the
closure density $(\Omega_{m} =3D 1)$ may eventually be reached; in this case=
,
the dark matter component
on large scales is distributed more diffusely than the luminous galaxies.

In a recent analysis, Bahcall, Lubin, and Dorman
[\cite{Bahcall:Lubin:1995}], show that this widely
believed scenario of an ever-increasing $M/L$ is not implied by the
available observations.

Bahcall, {\em et al.} [\cite{Bahcall:Lubin:1995}], examine the observed
$M/L$ ratio as a function of
scale for spiral and elliptical galaxies, groups, rich clusters, and
superclusters of galaxies. The
$M/L$ of spirals and ellipticals are determined from the virial mass based
on the velocities of stars
or gas in the galaxies within a given radius [\cite{FG:1979}];
[\cite{Bosma:1981}];
[\cite{Lauer:1985}];
[\cite{Trimble:1987}]; and references therein). To extend the estimated
galaxy mass beyond the luminous
or gas-dominated regimes, the authors use mass estimates derived from
motions of faint satellites at
larger distances around galaxies [\cite{Bahcall:1981}];
[\cite{Zaritzky:1993}], thus probing the dark
galaxy halos. They also use the extended X-ray halos observed around nearly
isolated ellipticals to
estimate their extended mass assuming that the X-ray gas is gravitationally
bound to the galaxy
[\cite{FG:1979}]; [\cite{Forman:1985}]; [\cite{Trimble:1987}];
[\cite{Fabbiano:1989}];
[\cite{Mushotzky:1994}].

Figure 1 presents the observed $M/L_{B}$ ratio as a function of radius for
a sample of bright spiral
and elliptical galaxies [\cite{Bahcall:Lubin:1995}]. A Hubble constant of
$H_{0} =3D 100 h$ km s$^{-1}$
Mpc$^{-1}$ is used and all luminosities, $L_{B}$, refer to the {\it total}
luminosity of the
galaxies in the blue band, corrected for Galactic and internal extinction
and redshift. The lines
connecting the data points indicate measurements at different radii {\it of
the same galaxy}. The
large boxes represent the $\sim 1 \sigma$ range of the observed $M/L_{B}$
at the Holmberg radius for a
large sample of bright ellipticals and spirals. As expected, $M/L_{B}$
increases with radius to scales
beyond the luminous part of the galaxies (the latter corresponding
typically to $R \lesssim 20$ kpc);
this increase reflects the well-established inference of {\it dark galaxy
halos}. The maximum extent
observed so far for an individual galaxy halo is $R \sim $ 200 kpc
[\cite{Bahcall:1981}];
[\cite{Kulessa:1992}];
[\cite{Zaritzky:1993}]. Most of the elliptical galaxy data on large scales
$(\gtrsim 20$ kpc) is
based on the extended X-ray halos observed around these galaxies.

The functions $M/L_{B} (R)$ that best fit the observations of spirals and
ellipticals are presented in
Figure 1. The average $M/L_{B}$ ratio of ellipticals appears to be larger
than that of spirals at the
same radius by a factor of approximately 3 or 4. (This is at least partly
due to the absence of bright
young blue stars in the old ellipticals). On average, thus, ellipticals
have more mass than spirals
for the {\it same} $L_{B}$ and radius. If the dark halos around typical
bright galaxies
extends to $R \sim 150-200$ kpc, as suggested by Figure 1, the implied
total $M/L_{B}$ ratio of the
galaxies is $M/L_{B} (\lesssim 0.2$Mpc) $\simeq 100$h for spirals and $\sim
400$h for ellipticals.
The large value of $M/L_{B}$ implied for ellipticals can thus contribute sig=
nificantly to the high
value of $M/L_{B}$ observed in rich clusters (see bleow) since clusters are
preferentially populated by
early type galaxies.

These results suggest that individual bright galaxies $(\sim L^{\star})$
are surrounded by very large
dark halos, extending typically to radii of $R \sim 150 - 200$ kpc. The
existence of such large
massive galactic halos is consistent with more recent observations of
gravitational lens distortions
of background objects by foreground galaxies [\cite{Brainerd:1996}] which
yield comparable radii
and masses for the dark galactic halos. The positive associations found
between Ly$\alpha$ clouds and
nearby galaxies up to radii of $R \sim$ 150 kpc [\cite{Barcons:1995}] is
also consistent
with the existence of dark galactic halos to these scales.

To compare the $M/L_{B}$ ratio of single galaxies with that of larger
systems such as groups and
rich clusters, Bahcall {\em et al.} [\cite{Bahcall:Lubin:1995}] used
optical and X-ray mass
determinations of groups and clusters of galaxies.

Figure 2 represents a composite of $M/L_{B}$ as a function of scale for
individual galaxies, groups,
and rich clusters. The data points represent median values of large
samples, as well as some
individual cases. Also included for comparison is the independent
$\Omega_{m}$ determination from the
cosmic virial theorem [\cite{Davis:1983}], utilizing galaxy pairwide
velocities at $r \lesssim 1
h^{-1}$ Mpc; approximate upper limits to $M/L_{B}$ for two large
superclusters; the {\it range} of
$\Omega_{m}$-values obtained from the  Virgocentric infall data, assuming
that mass follows light; the
$M/L_{B}$ ratio obtained by Tully, Shaya, and Peebles [\cite{Tully:1994}]=9F=
=9F
at $30 h^{-1}$Mpc using
the least action method; and the {\it range} of various recent reported
constraints of the $\beta =3D
\Omega_{m}^{0.6}/b$ parameter obtained from observations of bulk-flows and
redshift-space anisotropies
at $\sim 50 h^{-1}$ Mpc (for a bias parameter $b =3D 1$, i.e., mass traces l=
ight).

The $M/L_{B}$ ratio in Figure 2 increases with scale up to the largest
observed extent of individual
galaxies, $R \sim $ 200 kpc. Beyond this scale, however, $M/L_{B}$ appears
to ``flatten'' at an
intermediate value between spirals and ellipticals rather than increase
with scale; the observed
$M/L_{B}$ of groups and clusters typically range between $M/L_{B} \sim
100h$ to $400h$, as expected for
a mix of spirals ($\sim 100 h)$ and ellipticals ($\sim 400 h)$.

These observations suggest that the total mass of groups and clusters can
be accounted for by their
member galaxies plus the hot intracluster gas (observed to account for
$\sim 5\% - 10\% h^{-1.5}$ of
the total mass). The extended dark halos of galaxies may be stripped off in
dense cluster environments
but still remain in the clusters.  Since groups, clusters, and
superclusters are all composed of a mix
of spiral and elliptical galaxies, we expect that their $M/L_{B}$ (for a
system size $R \gtrsim 0.2
h^{-1}$ Mpc) will typically range from $\sim 100 h$ (if most members are
spirals) to $\sim 400 h$ (if
most are ellipticals). Under this scenario, the high $M/L_{B}$ of rich
clusters is mainly caused by the
high fraction of ellipticals in the clusters.

On larger scales, the supercluster data is less certain but also suggests
that the $M/L_{B}$ ratio does
not conitnue to rise significantly to large scales; rather, a constant
asymptotic value that is
consistent with a mixture of spirals and ellipticals may be suggested. The
bulk-flow and anisotropy
$\beta$ determinations on very large scale are too uncertain at the present
time to help constrain the
extension to these scales.

The flattening suggested in $M/L_{B}(R)$ (Fig. 2) implies that most of the
dark matter resides in
large galaxy halos with $R \sim $ 200 kpc. The mass of groups, clusters,
and possibly superclusters
of galaxies may be accounted for by the mass of their member galaxies,
including their large halos
(which may be stripped off in clusters but remain in the system), and the
mass of the observed
intracluster gas. No additional dark matter is needed to account for the=
 mass of these large systems.

\section{How Much Dark Matter?}

The optical and X-ray observations of rich clusters of galaxies yield
cluster masses that range from
$\sim 10^{14}$ to $\sim 10^{15}h^{-1} \; M_{0}$ within $1.5 h^{-1}$ Mpc
radius of the cluster center.
When normalized by the cluster luminosity, a median value of $M/L_{B}
\simeq 300 h$ is observed for
rich clusters. Gravitational lensing observations of weak distortions of
background galaxies by
foreground clusters yield cluster masses comparable to those obtained with
the optical and X-ray data.
The lensing results also suggest that the mass distribution in the clusters
follow approximately the
light distribution [\cite{Tyson:1996}]. These masses, and the mean
mass-to-light ratio, imply a
dynamical mass density of $\Omega_{\rm{dyn}} \sim 0.2$ on $1.5h^{-1}$ Mpc
scale.  If, however, the
universe has a critical density $(\Omega_{m} =3D 1)$, then most of the mass
in the universe {\it cannot}
be concentrated in clusters, groups, and galaxies; the mass would have to
be distributed more
diffusely than the light.

The analysis of the mass-to-light ratio of galaxies, groups and clusters
presented in Section 2
suggests that while the $M/L$ of galaxies increases with scale up to radii
of $R \sim 0.15 - 0.2
h^{-1}$ Mpc, due to the large dark halos around galaxies, this ratio
appears to flatten and remain
approximately constant for groups and rich clusters to scales of $\sim 1.5$
Mpc, and possibly even to
the larger scale of superclusters (Fig. 2). The flattening occurs at
$M/L_{B} \simeq 200-300h$,
corresponding to $\Omega_{m} \sim 0.2$. This observation suggests that most
of the dark matter is
associated with the dark halos of galaxies and that clusters do {\it not}
contain a substantial amount
of additional dark matter. If the $M/L_{B}(R)$ function indeed remains flat
to large scales, that
would suggest that the mass density of the universe is low, $\Omega_{m}
\sim 0.2$ (or $\Omega_{m} \sim
0.3$ for a small bias of $b \sim 1.5$).

Clusters of galaxies contain many baryons. Within $1.5h^{-1}$ Mpc of a rich
cluster, the X-ray emitting
gas contributes $\sim 5-10h^{-1.5}$\% of the cluster virial mass
[\cite{White:1995}];[\cite{Lubin:1996}]. Visible stars contribute only a
small additional amount to
this value. Standard Big-Bang nucleosynthesis limits the mean baryon
density of the universe to
$\Omega_{b} \sim 0.015h^{-2}$ [\cite{Walker:1991}]. This suggests that the
baryon fraction in some
rich clusters exceeds that of an $\Omega_{m} =3D 1$ universe by a large
factor [\cite{White:1993}];
[\cite{Lubin:1996}]. Detailed hydrodynamic simulations indicate that
baryons are not preferentially
segregated into rich clusters. It is therefore likely that either the mean
density of the universe is
considerably smaller, by a factor of $\sim 3$, than the critical density,
or that the baryon density
of the universe is much larger than predicted by nucleosynthesis. The
observed baryonic mass fraction
in rich clusters, when combined with the nucleosynthesis limit, suggest
$\Omega_{m} \sim 0.2-0.3$;
this estimate is consistent with $\Omega_{\rm{dyn}} \sim 0.2$ determined
from clusters. Visible matter
(stars and gas) contributes only a small fraction ($\sim$ 20\%) of the
total matter density (see
below).  This implies that even for $\Omega_{m} \sim 0.2$, most of the
matter is dark.

The above described methods of determining the mass denisty $\Omega_{m}$
are of a general nature,
independent of any cosmological model. They suggest a low value of
$\Omega_{m} \sim 0.2-0.3$. These
estimates are consistent with the values obtained from various large-scale
structure observations (for
a CDM cosmology; e.g., [\cite{Bahcall:Cen:1992}];
[\cite{Ostriker:1993}];[\cite{White:1993a}]. A low
value of $\Omega_{m}$ also reconciles the long age of the oldest stars with
the moderately high value
recently observed for the Hubble constant $(h \sim 0.6-0.7)$
[\cite{Freedman:1994}]; also Freedman,
this volume; Turner, this volume; but see Tammann, this volume).

\section{Nature of the Dark Matter}

We have seen in the previous section that most of the currently available
data suggest
that the mass-density of the universe may be low, $\Omega_{m} \simeq
0.2-0.3$. While this result is
not conclusive at the present time, observations over the next several
years, especially of
gravitational lensing and motions on large scales, should reveal a more
conclusive result.

How much of the estimated $\Omega_{m} \sim 0.2$ is baryonic? How much is
non-baryonic? Let us first
estimate what is the fraction of matter that exists as visible baryons,
i.e., gas and stars. The best
laboratory for observing baryons are rich galaxy clusters, which contain
substantial amounts of hot
intracluster gas detected through its X-ray emission.

The fraction of total cluster mass $(M_{cl})$ that is in the form of hot
gas is $M_{\rm{gas}}/M_{cl}
\simeq 20$\% (for $h \simeq 0.5$), and the fraction of stellar mass in
$M_{\star}/M_{cl} \simeq 5$\% ,
yielding a lower limit to the baryon fraction in clusters of $M_{b}/M_{cl}
\simeq
\Omega_{b} \Omega_{m} \gtrsim 25$\% (or $\sim$ 17\% for $h$ =3D 0.7). This
value is a lower limit since
only the ``visible'' baryons (in gas and stars) are counted; it is possible
that some or all of the
remaining matter in the cluster may be baryonic as well. If this fraction
is representative, as
expected, it would imply that $\gtrsim $ 20\% of the mass-density of the
universe is baryonic (for $h
\sim 0.5-0.7$). The constraints placed by nucleosynthesis, $\Omega_{b} \sim
0.06 h^{-2}_{50}$, combined
with the above observed ratio of $\Omega_{b}/\Omega_{m} \sim 0.25$, further
suggests that $\Omega_{m}
\simeq 0.2$, as discussed in Section 3.

What are some possible dark matter candidates?

\noindent {\underline {Baryonic Dark Matter}}. It is likely that some of
the dark matter is
baryonic. Some baryonic dark matter candidates include compact objects such
as MACHOS (Massive Compact
Halo Objects), which are currently being searched for (and several
candidates detected) by
microlensing searches (see Bennett, this volume). The current microlensing
candidates are reported to
be in the mass range of white-dwarfs.  Browns dwarfs do not appear to
provide a significant
contribution to the halo dark matter component. Baryonic dark matter in the
form of gas may also still
exist, although strong limits are placed on the amount of intergalactic HI
($\Omega_{HI} < 10^{-8}$)
and HII ($\Omega_{HII} \lesssim 0.04 h_{50}^{-2}$) (See [\cite{Carr:1994}].)

Could the dark matter be all baryonic? Not likely, due to the constraints
placed by nucleosynthesis
($\Omega_{b} \simeq 0.06 h_{50}^{-2}$) as compared with the observed
$\Omega_{m} \sim 0.2$. In
addition, observational searches for dark matter candidates are narrowing
the allowed range of
possibilities [\cite{Carr:1994}]; it has not been easy to find candidate
dark matter objects!
\vspace{.2in}

\noindent {\underline{Non-Baryonic Dark Matter}}

The main types of non-baryonic dark matter candidates considered are cold
(CDM) and hot (HDM). The CDM
candidates (such as axions), provide the needed seeds for galaxy formation,
and fit well (for low
$\Omega_{m}$) various large scale structure observations (including the
power spectrum and correlation
function of galaxies, and the cluster mass function and correlation
function). The main``problem'' for
CDM is that such particles are not yet known to exist (!). In addition, the
standard, more elegant
$\Omega_{m} =3D 1$ CDM model does not fit the available data. A more ad-hoc
and less elegant $\Omega_{m}
\sim 0.2-0.3$ CDM model, or other variants of CDM, are needed in order to
satisfy current
large-scale structure observations.

\noindent The HDM candidates, such as massive neutrinos, are unlikely to
comprise a major part of the
dark matter component since the hot matter inhibits galaxy formation at
high redshifts, in
contradiction with observations.

\section{Conclusions}

Observations suggest some possible clues to the questions introduced in
Section 1; I sumarize the
conclusions below.

\begin{itemize}

\item How much dark matter=20is there in the Universe?

Most of the reliable observations at present
suggest that $\Omega_{m} \sim 0.2 - 0.3$. The amount of visible (luminous)
matter, in optical and
X-ray light, is only a small fraction of the above total ($\sim$ 20\% for
$h \sim 0.5 - 0.7$). Most of
the matter is therefore dark.

\item Where is the dark matter located?

Observations indicate that most of the dark matter is located in very large
galactic halos, extending to radii of $\sim 200 $ kpc
[\cite{Bahcall:Lubin:1995}]. The galaxies and
their halos also comprise the main mass of groups and clusters of galaxies
(where some of the halos
may be striped off but still reside in the clusters). No significant
additional dark matter is
needed to account for the mass of groups and clusters.

\item What is the nature of the dark matter?

\begin{itemize}

\item Baryons comprise at least 20\% of the matter (for $h \sim 0.7$) .

\item Cold dark matter may possibly exist; it provides a good fit to the
observed large scale
structure of the Universe, and produces necessary seeds for galaxy formation=
s.

\item Hot dark matter may exist, but cannot be the dominant part of the
dark matter since it hinders
galaxy formation at high redshifts.

\item Baryonic dark matter is very likely part of the dark matter
component. Compact objects such as
MACHOs are among the likely candidates.  We note however, that if the
entire halos to $R \sim 200$ kpc
are dominated by baryonic compact objects, the baryon density would exceed
that predicted by big-bang
nucleosynthesis.
\end{itemize}
\end{itemize}

It is a pleasure to thank the organizers of this outstanding conference in
honor of Princeton
University's 250$^{\rm{th}}$ anniversary celebration, J. Gunn, J. Ostriker,
J. Peebles, D. Spergel, N.
Turok, and D. Wilkinson for a most stimulating, fruitful and lively
conference. The work by N. Bahcall
and collaborators is supported by NSF Grant AST-93-15368 and NASA Grant
NGT-51295.

\vfill\eject

\begin{thebibliography}{}
\bibitem{FG:1979} [1]  Faber, S.M., \& Gallagher, J.S., 1979, {\em ARA\&A},
{\bf 17}, 135.\\
\bibitem{Trimble:1987}  [2] Trimble, V., 1987, {\em ARA\&A}, {\bf 25}, 423.\=
\
\bibitem{Peebles:1993}  [3] Peebles, P.J.E., 1993, {\em Principles of
Physical Cosmology}, (Princeton:
Princeton University Press).  \\
\bibitem{Ostriker:1974}  [4] Ostriker, J.P., Peebles, P.J.E., and Yahil,
A., 1974, {\em Ap.J.}, {\bf
193} L1.\\
\bibitem{Einasto:1974}  [5] Einasto, J., Kaasik, A., and Saar, E., 1974.
{\em Nature}, {\bf 250},
309.\\
\bibitem{Davis:1980} [6] Davis, M., Tonry, J., Huchra, J., and Latham, D.,
1980, {\em Ap.J.}, {\bf
238}, L113.\\
\bibitem{Lauer:1985}[7] Lauer, T.R., 1985, {\em Ap.J.}, {\bf 292}, 104.\\
\bibitem{Forman:1985}[8] Forman, W., Jones, C., and Tucker, W., 1985, {\em
Ap.J.}, {\bf 293}, 102.\\
\bibitem{Fabbiano:1989}[9] Fabbiano, G., 1989, {\em ARA\&A}, {\bf 27}, 87.\\
\bibitem{Mushotzky:1994}[10] Mushotzky, R.F., Loewenstein, H., Awaki, K.,
Makishima, K., Matsushita,
K., and Matsumoto, H., 1994, {\em Ap.J.}, {\bf 436}, L79.\\
\bibitem{Lubin:1993} [11] Lubin, L.M., and Bahcall, N.A., 1993, {\em
Ap.J.}, {\bf 415}, L17.\\
\bibitem{Kaiser:1993}[12] Kaiser, N., and Squires, G., 1993, {\em Ap.J.},
{\bf 404}, 441.\\
\bibitem{Tyson:1996} [13] Tyson, A., and Fischer, D., 1996, {\em Ap.J.}, in
press.\\
\bibitem{Brainerd:1996} [14] Brainerd, T., Blanford, R., and Smail, I.,
1996, {\em Ap.J.}, {\bf 466},
623.\\
\bibitem{Blumenthal:1984}[15] Blumenthal, G.R., Faber, S., Primack, J.R.,
and Rees, M.J., 1984, {\em
Nature}, {\bf 311}, 517. \\
\bibitem{Rubin:1993}[16] Rubin, V.C., 1993, in {\em Proceedings Natl. Acad.
Sci.}, {\bf 90}, 4814.\\
\bibitem{Bahcall:Lubin:1995}[17] Bahcall, N.A., Lubin, L.M., and Dorman,
V., 1995, {\em Ap.J.}, {\bf
447}, L81.\\
\bibitem{Bosma:1981} [18] Bosma, A., 1981, {\em AJ}, {\bf 86}, 1825.\\
\bibitem{Bahcall:1981} [19] Bahcall, N.A., and Tremaine, S., 1981, {\em
Ap.J.}, {\bf 244}, 805.\\
\bibitem{Zaritzky:1993} [20] Zaritzky, D., Smith, R., Frenk, C. and White,
S.D.M., 1993, {\em Ap.J.},
{\bf 405}, 464.\\
\bibitem{Kulessa:1992} [21] Kulessa, A.S., and Lynden-Bell, D., 1992, {\em=
 MNRAS}, {\bf 255}, 105.\\
\bibitem{Barcons:1995}[22]  Barcons, X., Lanzetta, K.M., and Webb, J.K.,
1995, {\em Nature}, {\bf 376},
321.\\
\bibitem{Davis:1983}[23] Davis, M., and Peebles, P.J.E., 1983, {\em Ap.J.},
{\bf 267}, 465.\\
\bibitem{Tully:1994} [24] Tully, R.B., Shaya, E.J., and Peebles, P.J.E.,
1994, in {\em Proc. Yamada
Conference 23}, ed. K. Sato (Tokyo: Universal Academy Press), 217.\\
\bibitem{White:1995}[25] White, D. and Fabian, A., 1995, {\em MNRAS}, {\bf
273}, 72.\\
\bibitem{Lubin:1996}[26] Lubin, L.M., Cen, R., Bahcall, N.A., and Ostriker,
J.P., 1996, {\em Ap.J.},
{\bf 460}, 10.\\
\bibitem{Walker:1991}[27] Walker, T.P., {\em et al.}, 1991, {\em Ap.J.},
{\bf 376}, 51.\\
\bibitem{White:1993}[28] White, S.D.M., Navarro, J.F., Evrard, A.E., and
=46renk, C.S., 1993, {\em
Ap.J.}, {\bf 366}, 429.\\
\bibitem{Bahcall:Cen:1992} [29] Bahcall, N.A., and Cen, R., 1992, {\em
Ap.J.}, {\bf 398}, L81.\\
\bibitem{Ostriker:1993} [30] Ostriker, J.P., 1993, {\em ARAA}, {\bf 31}, 689=
.\\
\bibitem{White:1993a} [31] White, S.D.M., Efstathiou, G., and Frenk, C.S.,
1993, {\em MNRAS}, {\bf
262}, 1023.\\
\bibitem{Freedman:1994}[32] Freedman, W.L., {\em et al.}, 1994, {\em
Nature}, {\bf 371}, 757.\\
\bibitem{Carr:1994} [33] Carr, B., 1994, {\em MNRAS}, {\bf 32}, 531.\\
\end{thebibliography}
\end{document}